\documentclass[amsmath,amssymb,aps,pra,reprint,superscriptaddress,footnoteinbib]{revtex4-1} 
\usepackage{amsmath,amssymb}
\usepackage[english]{babel}
\usepackage{ bbold }
\usepackage{upgreek}
\usepackage{dsfont}
\usepackage{graphicx}
\usepackage{comment}
\usepackage{braket}
\usepackage{colortbl}
\usepackage{blindtext}
\usepackage{hyperref}
\begin{document}

\title{Torque-free manipulation of nanoparticle rotations via embedded spins}

\author{Yue Ma}
\affiliation{QOLS, Blackett Laboratory, Imperial College London, London SW7 2AZ, United Kingdom}

\author{M. S. Kim}
\affiliation{QOLS, Blackett Laboratory, Imperial College London, London SW7 2AZ, United Kingdom}

\author{Benjamin A. Stickler}
\affiliation{QOLS, Blackett Laboratory, Imperial College London, London SW7 2AZ, United Kingdom}
\affiliation{Faculty of Physics, University of Duisburg-Essen, Lotharstra\ss e 1, 47048 Duisburg, Germany}

\begin{abstract}
Spin angular momentum and mechanical rotation both contribute to the total angular momentum of rigid bodies, leading to spin-rotational coupling via the Einstein-de Haas and Barnett effects. Here we show that the revolutions of symmetric nanorotors can be strongly affected by a small number of intrinsic spins. The resulting dynamics are observable with freely rotating nanodiamonds with embedded nitrogen-vacancy centers and persist for realistically-shaped near-symmetric particles, opening the door to torque-free schemes to control their rotations at the quantum level.
\end{abstract}

\maketitle

\section{Introduction}

Levitating nanoscale particles in ultrahigh vacuum provides a promising platform for high-mass tests of quantum physics and for ultra-precise force and torque sensors \cite{millen2020optomechanics,millen2020quantum}. State-of-the-art experiments with levitated objects include cooling their center-of-mass to the quantum ground state \cite{delic2020,magrini2020optimal,tebbenjohanns2021quantum}, controlling their alignment~\cite{hoang2016,kuhn2017,kuhn2017part2,rashid2018,ahn2020,vanderlaan2020}, and spinning them with GHz frequencies~\cite{ahn2018,reimann2018,jin20206}. Rotational cooling has been achieved recently~\cite{delord2020,bang2020,van2020observation}, with the quantum regime within reach \cite{schafer2021cooling}.  The inherent non-linearity of rigid body rotations  \cite{landau1976mechanics,stickler2021quantum} gives rise to pronounced quantum interference phenomena \cite{stickler2018,ma2020}, rendering the rotational degrees of freedom particularly attractive for future quantum applications.

The rotational motion of nanoscale particles is modified by embedded spins \cite{rusconi2016magnetic,rusconi2017quantum} due to the Einstein-de Haas \cite{einstein1915experimental} and Barnett effects~\cite{barnett1915magnetization}, which express that spin and mechanical angular momentum can be interconverted \cite{chudnovsky1994conservation,kovalev2005nanomechanical,garanin2011quantum,ganzhorn2016quantum,dornes2019ultrafast,mentink2019quantum}. While spin-rotational coupling is of minor importance for the rotations of macroscopic objects, it opens the door to new strategies for controlling and detecting nano- to microscale magnetized rotors  \cite{pratcamps2017ultrasensitive,wang2019dynamics,gieseler2020single,vinante2020ultralow}. Thus far, spin-rotational coupling has been explored for magnets whose spin angular momentum dominates over the mechanical one~\cite{kimball2016precessing,fadeev2021ferromagnetic,vinante2021surpassing}, enabling for instance the levitation of non-rotating spherical nanomagnets despite Earnshaw's theorem \cite{rusconi2017quantum,rusconi2017linear,kustura2021quantum}.

\begin{figure}
\centering
\includegraphics[width=0.48\textwidth]{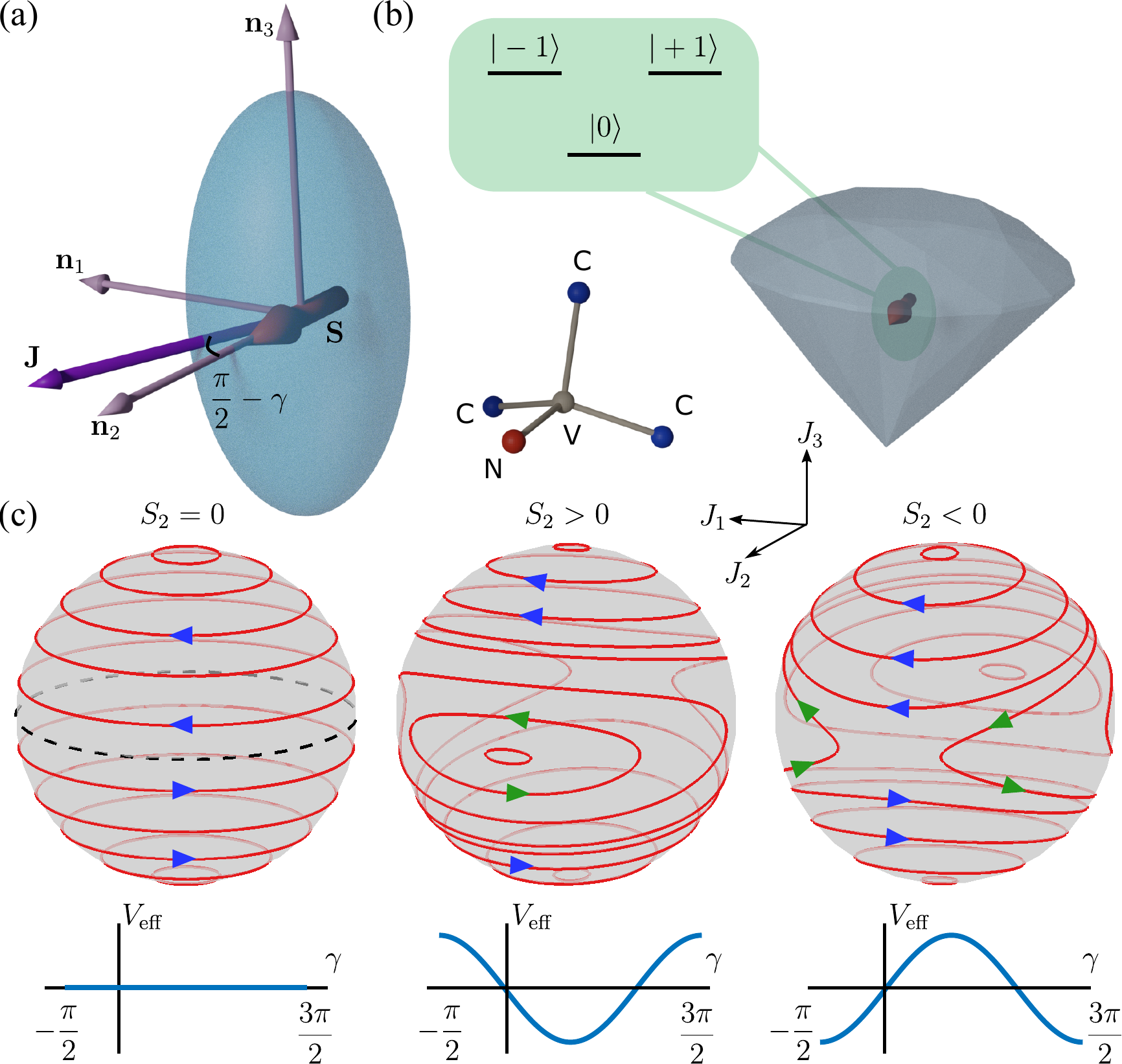}
\caption{The torque-free rotation of a prolate rotor can be strongly modified by embedded spins. (a) The particle rotates rapidly around an axis approximately orthogonal to its symmetry axis ${\bf n}_3$ while its embedded spin (magenta vector) is aligned with the body-fixed ${\bf n}_{2}$ axis. The conservation of the total angular momentum vector (purple arrow) and the Barnett/Einstein-de Haas effect can strongly modify the particle rotations {\it around} the symmetry axis, as described by the angle $\gamma$. (b) The spin can for instance be due to nitrogen vacancy (NV) centers in a nanodiamond, whose energy levels and simplified lattice structure are shown in the insets. (c) The presence of the spin modifies the body-fixed angular momentum trajectories of the free rotor ($S_2 = 0$) by adding oscillatory trajectories close to the equator. Depending on the sign of $S_2 = {\bf n}_2 \cdot {\bf S}$, the spin can trap the rotor trajectories close to $J_2 = \pm J $ for $S_2 \gtrless 0$, respectively. The resulting dynamics of the spin axis ${\bf n}_2$ are governed by an effective potential (bottom row) of magnitude $S_2J/I$.}\label{fig:model}
\end{figure}

In this paper we show that only a few embedded spins can strongly modify the torque-free dynamics of symmetric nanorotors. This is a result of the non-linearity of rotations and of the conservation of total angular momentum. As a concrete example, this can lead to stabilization or destabilization of the nanoparticle rotation depending on the relative alignment of spin and mechanical angular momentum. This strong spin-rotational coupling is observable with realistically-shaped near-symmetric nanodiamonds containing nitrogen-vacancy (NV) centers.

Nanodiamonds with embedded NVs have been proposed to enable superposition tests \cite{scala2013,yin2013large,wan2016,kumar2017magnetometry,ma2017proposal,delord2017strong,delord2017electron,conangla2018motion,delord2018ramsey,pedernales2020motional,huillery2020spin,perdriat2021spin}, tests of the quantum nature of gravity \cite{bose2017spin}, and might allow high-precision rotation sensing \cite{maclaurin2012measurable,wood2017magnetic,wood2018quantum,wood2020observation,chen2019nonadiabatic,zangara2019mechanical}. We discuss how strong spin-rotational coupling can be exploited in future rotational quantum experiments with nanodiamonds. As a first signature of quantum coherence, we study the breakdown of the hard-magnet regime at GHz rotation rates due to rotation-induced spin transitions.

\section{Torque-free rigid body rotation}

We consider the rotation of a rigid body, characterized by its inertia tensor ${\rm I}$ with moments of inertia $I_{1,2,3}$ and principal axes $\mathbf{n}_{1,2,3}$. In the absence of external torques, the total angular momentum $\mathbf{J}= {\rm I}\boldsymbol{\omega}+\mathbf{S}$ is conserved, containing both the mechanical angular momentum with velocity vector $\boldsymbol{\omega}$, defined by $d{\bf n}_k/dt = \boldsymbol{\omega}\times {\bf n}_k$, and the embedded spin angular momentum $\mathbf{S}$. The latter can be related to the magnetic moment $\mathbf{m}$ by $\mathbf{S}=-\hbar\mathbf{m}/g_s\mu_B$, where $g_s$ is a material specific factor and $\mu_B$ is the Bohr magneton. If the body-fixed components  $S_k =  \mathbf{S}\cdot\mathbf{n}_k$ are constant, so that the spin ${\bf S}$ rotates together with the particle (hard-magnet regime), the nonlinear Euler equations for the body-fixed components $J_k= \mathbf{J}\cdot\mathbf{n}_k$ are
\begin{equation}
    \frac{d}{dt}J_i=\frac{I_j-I_k}{I_kI_j}J_jJ_k-\frac{1}{I_k}S_kJ_j+\frac{1}{I_j}S_jJ_k,\label{eq:euler}
\end{equation}
where $(i,j,k)$ is an even permutation of $(1,2,3)$. Spin-rotational coupling via the Einstein-de Haas and Barnett effects \cite{einstein1915experimental,barnett1915magnetization} is described by the terms containing the components of $\mathbf{S}$. We will next show how these terms can give rise to strong spin-rotational coupling despite $S_k \ll J = |{\bf J}|$. 

\section{Symmetric rotor}

We first study a symmetric rotor with $I_1=I_2 = I \neq I_3$ rotating rapidly around an axis approximately orthogonal to the symmetry axis ${\bf n}_3$ and containing an embedded spin aligned with the body-fixed ${\bf n}_2$ axis, i.e. $S_1 = S_3 = 0$ with $|S_2|\ll J$ [see Fig.~\ref{fig:model}(a)]. In this case the spin can strongly affect the particle rotations. This can be seen by choosing the space-fixed axis so that ${\bf J}=  J{\bf e}_z$, implying that the rotor symmetry axis approximately revolves with constant speed $J/I$ in the space-fixed $x$-$y$ plane (gyroscopic stabilization). The rotations {\it around} the symmetry axis, with angle $\gamma$, follow from the Euler equations \eqref{eq:euler} by using that $J_1 \simeq - J \cos \gamma$, $J_2 \simeq J \sin \gamma$ and $J_3 \simeq I_{\rm eff}  \dot{\gamma} \ll J$, where we abbreviated the effective moment of inertia $I_{\rm eff} = I I_3/(I - I_3)$ \cite{goldstein,rusconi2016magnetic} (see Appendix~\ref{sec:Ham}). Substituting this into the Euler equation \eqref{eq:euler} shows that the dynamics are described by the effectively one-dimensional Hamiltonian
\begin{equation}\label{eq:heff}
    H_{\rm eff} = \frac{p_\gamma^2}{2I_{\rm eff}} - \frac{S_2 J}{I}\sin \gamma,
\end{equation}
where $p_\gamma$ is the angular momentum for rotations around the symmetry axis, so that $p_\gamma = I_{\rm eff} \dot{\gamma}$  (see Appendix~\ref{sec:Ham}). The spin thus adds an effective potential of magnitude $S_2J/I$ to the rotations around the symmetry axis of the rotor.

The Hamiltonian Eq.~\eqref{eq:heff} describes how the spin affects the rotations around the symmetry axis. The spin-induced effective potential vanishes for $S_2 = 0$, meaning that the principal axes ${\bf n}_1$ and ${\bf n}_2$ rotate uniformly with constant angular velocity $p_\gamma/I_{\rm eff}$ around the symmetry axis ${\bf n}_3$. For finite spin, the dynamics is similar to that of a physical pendulum in linear gravity, that is the product of spin $S_2$ and total angular momentum $J$ generate an anharmonic potential acting on the motion of the angle $\gamma$. The physical origin of this effective potential is the Barnett/Einstein-de Haas effect: the fast rotation of the particle acts on the spin as a synthetic magnetic field proportional to $J/I$, forcing the spin to align with the total angular momentum vector ${\bf J}$. Since the spin is rigidly attached to the particle (hard-magnet regime), this acts back on the rotation of the particle. Whether the spin axis tends towards the total angular momentum vector or its opposite direction depends on whether the rotor is prolate ($I > I_3$), resulting in a positive effective moment of inertia, or oblate ($I < I_3$), giving rise to a negative effective moment of inertia. Fig.~\ref{fig:model}(c) shows the resulting angular momentum trajectories for a prolate rotor for three different $S_2$.

The Hamiltonian \eqref{eq:heff} implies that the spin axis ${\bf n}_2$ in a prolate rotor will remain close to its initial orientation $\gamma(0) \simeq \pi/2$ if
\begin{equation}
    S_2\gtrsim\frac{5}{2}\left ( \frac{I}{I_3} -1 \right )\frac{p_\gamma^2(0)}{J},\label{eq:sym}
\end{equation}
resulting in $\sin \gamma \geq 4/5$ for all times. This spin-induced stabilization of rotations around the symmetry axis requires no external torques and works even though $S_2 \ll J$. Likewise, if $S_2<0$ the spin axis  will be {\it destabilized} when initially oriented at $\gamma(0) \simeq \pi/2$. Importantly, this strong influence of spin on the rotation dynamics persists for realistically-shaped objects, as we will show next.

\section{Near-symmetric rotor}

Asymmetric rotors exhibit the mid-axis instability \cite{goldstein}, rendering rotations around the axis of intermediate moment of inertia dynamically unstable. Thus even in the absence of embedded spins, the angle $\gamma$ experiences an effective potential. For a near-prolate asymmetric rotor with $I_1 \gtrsim I_2 > I_3$, a straightforward calculation shows that this adds the term $(I_1 - I_2) J^2 \sin^2 \gamma/2I_1I_2$ (see Appendix~\ref{sec:Ham}) to the Hamiltonian \eqref{eq:heff}. This implies that the spin converts the effective potential to a trapping potential around $\gamma\simeq\pi/2$ if
\begin{equation}
    S_2\gtrsim\frac{I_1-I_2}{I_1}J,\label{eq:asymS}
\end{equation}
while simultaneously enhancing the inverse-trapping potential around $\gamma\simeq 3\pi/2$. This relation in combination with Eq.\,\eqref{eq:sym} demonstrates under which conditions the mid-axis rotations of a near-prolate object can be stabilized despite the instability. We will show next that this strong spin-rotational coupling is observable with nanodiamonds containing embedded NV centers, where  $|S_2| \ll J$ [see Fig.~\ref{fig:model}(b)].

The case of near-oblate rotors ($I_3>I_2 \gtrsim I_1$) follows straightforwardly by flipping the sign of the effective moment of inertia.

\begin{figure}[t]
\centering
\includegraphics[width=0.48\textwidth]{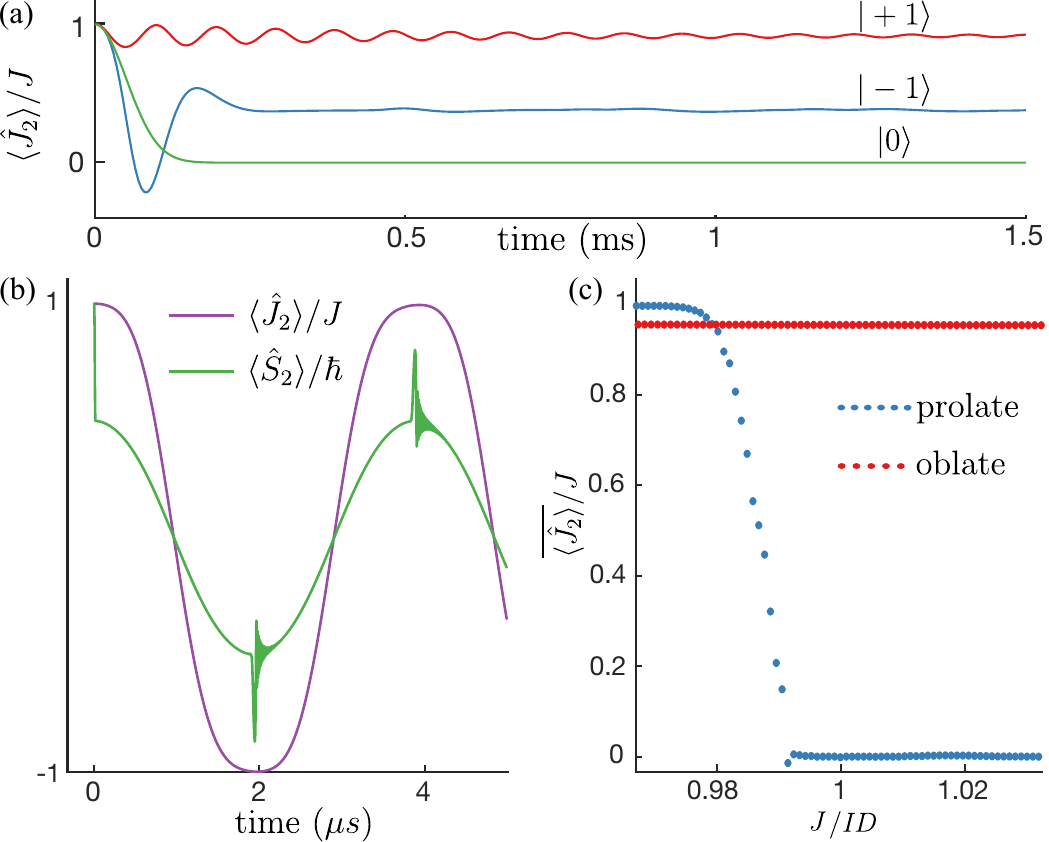}
\caption{Examples of how a single NV center modifies the rotation of a symmetric nanodiamond. (a) For a prolate particle rotating off resonance, $J/I \ll D$, the Gaussian decay of the alignment $\langle {\bf e}_2 \cdot {\bf n}_2\rangle\simeq \langle \hat{J}_2 \rangle/J$ (green) is suppressed by preparing the NV spin in the $|+1\rangle$ state (red), and can be converted into a different decay pattern by the $|-1\rangle$ state (blue). (b) The time evolution of the mean value of $\hat{J}_2$ (purple) and $\hat{S}_2$ (green) for a prolate particle rotating on resonance, shows nonadiabatic transitions when $|\langle \hat{J}_2 \rangle|\simeq J$ and adiabatic dynamics otherwise. (c) The effects of rotation-induced spin transitions are shown via the time-averaged mean angular momentum, $\overline{\langle \hat{J}_2\rangle}$. For on resonance rotation $J\simeq ID$, the stabilization of a prolate rotor by the $|+1\rangle$ state breaks down due to spin transitions, leading to a reduction of $\overline{\langle \hat{J}_2\rangle}$ (blue). An oblate rotor remains stabilized by the $|-1\rangle$ state (red) as spin transitions are suppressed.}\label{fig:diamond}
\end{figure}

\section{Nanodiamond with NV centers}

The Hamiltonian of a nanodiamond rotor with $N$ embedded NV centers is in the absence of external torques given by~\cite{rusconi2016magnetic,ma2017proposal}
\begin{equation}
\hat{H}_{\mathrm{tot}}=\sum_{k=1}^3\frac{1}{2I_k}\left (\hat{J}_k-\sum_{m=1}^N\hat{S}_k^{(m)}\right )^2+\frac{D}{\hbar}\sum_{m=1}^N\left(\mathbf{n}^{(m)}\cdot\hat{\mathbf{S}}^{(m)}\right)^2,
    \label{eq:quantum_H}
\end{equation}
where $D=2\pi\times2.87\ \mathrm{GHz}$ is the zero field splitting and $\mathbf{n}^{(m)}$ and $\hat{\mathbf{S}}^{(m)}$ are the quantization axis and spin operator of the $m$-th NV center. The distance between the NV centers is assumed to be large enough so that spin-spin interactions are negligible. The body-fixed total angular momentum operators $\hat{J}_{i}$ fulfill the anomalous commutation relations $[\hat{J}_i,\hat{J}_j]=-i\hbar\epsilon_{ijk}\hat{J}_k$~\cite{edmonds1996}, while the body-fixed spin angular momentum operators  $\hat{S}_{i}^{(m)}$ obey $[\hat{S}_i^{(m)},\hat{S}_j^{(n)}]=i\hbar\epsilon_{ijk}\delta_{mn}\hat{S}_k^{(m)}$~\cite{di2020rotational}. Note that $\hat{J}_i$ and $\hat{S}_j^{(m)}$ commute for every $i,j,m$  \cite{rusconi2016magnetic}.

If the mechanical rotation rate is much smaller than the NV zero-field splitting, $J/I \ll D$, the spin eigenstates of ${\bf n}^{(m)}\cdot \hat{\bf S}^{(m)}$ do not evolve (rotating wave approximation, see Appendix~\ref{sec:rwa}). The particle thus behaves as a hard magnet and the classical analysis of spin-rotational coupling applies. To demonstrate that strong spin-rotational coupling can be observed with nanodiamonds we consider a near-prolate particle whose NV center quantization axes are along the ${\bf n}_2$  axis, $\mathbf{n}^{(m)}\cdot \mathbf{n}_2\simeq 1$. The particle is initially trapped and rotationally cooled \cite{schafer2021cooling} to maximally align its body-fixed axes with the space-fixed frame. Once the particle is aligned, the spins are polarized by a microwave pulse into the desired spin state and a static magnetic field keeps the rotor in place while it is angularly accelerated to large angular momentum around ${\bf n}_2$. This creates the approximately thermal state~\cite{ma2020},
\begin{equation}
     \rho\simeq\frac{1}{Z}\exp\left[-\frac{1}{k_{\rm B} T}\left(\frac{\hat{J}_1^2}{2I_1}+\frac{(\hat{J}_2-J)^2}{2I_2}+\frac{\hat{J}_3^2}{2I_3}\right)\right],\label{eq:Tstate}
\end{equation}
where $Z$ is the partition function. The displacement $J$ describes the initial rapid rotation around the mid-axis and dominates the thermal width $J \gg \sqrt{I_k k_{\rm B} T}$. The nanodiamond is then released and rotates freely. The orientation of $\mathbf{n}_2$ can for instance be read out via the NV states~\cite{delord2017electron,delord2018ramsey}. For large quantum number (e.g. $J/\hbar>100$), the mean value of the relative orientation  $\langle {\bf e}_2 \cdot {\bf n}_2\rangle$ between body and space fixed axes is well approximated by the corresponding classical dynamics.

The condition Eq.~\eqref{eq:sym} for spin-rotational stabilization of $\mathbf{n}_2$ is washed out by the thermal state because the initial angular momentum $p_\gamma(0)$ becomes normally distributed. Estimating the initial angular momentum $p_\gamma^2(0)$ by half of the thermal width $I_3 k_{\rm B} T$ and neglecting prefactors of order unity, one obtains
\begin{equation}
    T\lesssim\frac{S_2J}{k_{\mathrm{B}}(I-I_3)},\label{eq:thermalC}
\end{equation}
where the value of $S_2$ is fixed by the initialization of the NV spins. Equation~\eqref{eq:thermalC} can be fulfilled even for $S_2$ as small as $\hbar$ from a single NV center for fast enough rotation.

As a numerical example [Fig.~\ref{fig:diamond}(a)], we consider the quantum dynamics of a prolate ellipsoidal nanodiamond with semiaxes $10$ nm and $11$ nm rotating around $\mathbf{n}_2$ (i.e. $\gamma(0) = \pi/2$) at frequency $\omega/2\pi=23.7\ \mathrm{MHz}$. (The choice of size is limited by numerical constraints.) At the experimentally achievable temperature $T=2\ \mathrm{mK}$ \cite{schafer2021cooling}, the rotor $\mathbf{n}_2$ axis is stabilized close to its initial orientation by a single NV center prepared in the $|+1\rangle$ eigenstate. In contrast, initializing the NV center in the $|0\rangle$ state leads to a Gaussian decay of the alignment $\langle {\bf e}_2 \cdot {\bf n}_2\rangle$ due to the thermal width in $\hat{J}_3$. The characteristic decay time follows from Eqs.\,\eqref{eq:euler} to be inversely proportional to the thermal width in $J_3$, $\tau_{\mathrm{sym}}=I_3I/(I-I_3)\sqrt{I_3k_{\rm B}T}$. Finally, if the NV spin is in the $|-1\rangle$ state, the rotor first quickly flips due to the destabilization, and then $\langle \mathbf{e}_2 \cdot \mathbf{n}_2 \rangle$ approaches a constant positive value due to the thermal width and dispersion of the rotor~\cite{landau1976mechanics}.

The degree of symmetry required to observe spin-induced stabilization is rather demanding for small nanodiamonds since $I_1 = I_2$ is technologically impossible. Equation \eqref{eq:asymS} implies that rod-shaped particles are preferable, as the length of the approximate symmetry axis contributes to both $I_1$ and $I_2$, masking small differences between the two short axes. Apart from this challenge, the rotor asymmetry also has benefits for observing the spin-induced stabilization. The mid-axis flipping of an asymmetric rotor happens on shorter timescales than for symmetric objects \cite{ma2020}, thus the effect of the spin is observable at earlier times. Requiring the spin angular momentum to satisfy \eqref{eq:asymS} also means that \eqref{eq:thermalC} can be reached at a smaller $J$ or at a higher temperatures, relaxing the requirements on rotational cooling. For instance, strong spin-rotational coupling can be observed with an ellipsoidal nanodiamond with semiaxes $10$ nm,$11$ nm and $200$ nm. At $T=0.1\ \mathrm{mK}$ and $J/I_2\simeq 2\pi\times9\ \mathrm{kHz}$, the free dispersion of $\mathbf{n}_2$ becomes significant on the time scale of $0.1$\,ms. Embedding as many as $800$ NV centers whose quantization axes are approximately parallel to $\mathbf{n}_2$ and initializing them to the state $|+1\rangle^{\otimes N}$, can stabilize $\mathbf{n}_2$. Likewise, initializing the spins in the $|-1\rangle^{\otimes N}$ state destabilizes the rotor. Equation~\eqref{eq:thermalC} is also the requirement for stabilizing near-oblate rotors ($I_3>I_2\gtrsim I_1$), implying $S_2<0$.

For small rotation rates, $J/I \ll D$, the particle remains in the hard-magnet regime, so that its rotation dynamics can be understood classically. However, signatures of quantum coherent spin transitions can emerge for faster rotations.

\section{Rotation on resonance}

If the particle rotation rate $J/I$ becomes comparable to the NV zero-field splitting the rotation can induce spin transitions, leading to a breakdown of the hard-magnet regime. One manifestation of these quantum coherent  spin transitions is that an initially stabilized rotor can turn unstable on relatively short timescales. To see this, we consider a near-prolate rotor initially rotating rapidly around $\mathbf{n}_2$, with NV centers whose quantization axes are along $\mathbf{n}_2$ and initialized to $|+1\rangle$, satisfying the hard-magnet regime stabilization condition \eqref{eq:thermalC}. The energy spacing between the $m$-th NV center in the $|+1^{(m)}\rangle$ state and in the $|0^{(m)}\rangle$ state becomes small, $\Delta E_{+1}=\hbar (D-J/I) \ll \hbar D$, and thus nonadiabatic transitions between these states are possible on short timescales. This breaks the condition \eqref{eq:thermalC} because $\langle S_2\rangle$ decreases, forcing $\langle J_2 \rangle$ to also decrease until the rotations are no longer on resonance with the spin-transitions, suppressing the latter. In contrast, a near-oblate rotor with $J_2(0)\simeq ID$ stays stabilized by preparing the NV in the $|-1\rangle$ state with quantization axis $\mathbf{n}_2$ because the energy spacing between the $|-1^{(m)}\rangle$ and the $|0^{(m)}\rangle$ states remains large, $\Delta E_{-1}\simeq 2\hbar D$, suppressing spin transitions.

Figures~\ref{fig:diamond}(b,c) illustrate the dynamics of a prolate and an oblate rotor containing one NV center whose quantization axis is parallel to $\mathbf{n}_2$. The semiaxes of the prolate (oblate) rotor are $6$ nm and $7$ nm ($6.81$ nm and $5.44$ nm), chosen such that their dynamics are equivalent in the hard-magnet regime. The temperature is set to zero in order to solely concentrate on the NV transitions. Fig.~\ref{fig:diamond}(b) shows the dynamics of the prolate rotor and its embedded NV center for $J=ID$. When $|\langle \hat{J}_2 \rangle|\simeq J$, nonadiabatic NV state transitions are manifested as a sharp change in $\langle \hat{S}_2 \rangle/\hbar$ while $\langle \hat{J}_2 \rangle/J$ remains almost constant. A qualitative semiclassical explanation of the dynamics combining classical rotation trajectories with quantum NV states is provided in Appendix~\ref{sec:onNV}. Fig.~\ref{fig:diamond}(c) plots the degree of stabilization as a function of the ratio $J/ID$ close to resonance. The degree of stabilization is quantified by the time averaged mean value $\overline{\langle \hat{J}_2\rangle} = \int_0^{T_0} dt \langle \hat{J}_2(t)\rangle/T_0$, where $T_0=150\ \mu$s is chosen on the order of NV center $T_1$ time in nanodiamond. The latter has been measured to be $\sim 100\ \mu$s at cryogenic temperatures and is expected to be longer for improved surface qualities~\cite{guillebon2020}. The rotation-induced spin transitions occur on much shorter timescales, so that the $T_2^*$ time poses no limit to the proposed experiment. The reduction of $\overline{\langle \hat{J}_2\rangle}$ towards $0$ for a prolate rotor indicates the breakdown of the stabilization while the oblate rotor remains stabilized.

\section{Conclusion}

In this paper we studied the impact of spin-rotational coupling on the three-dimensional rotation dynamics of nanorotors and identified a regime where the torque-free rotational dynamics of a near-symmetric rotor can be modified strongly by only a few embedded spins. This manifestation of the Einstein-de Haas and Barnett effects can be observed with objects that are not ferromagnets, such as nanodiamonds containing NV centers, as the spin-induced stabilization or destabilization of their rotations around the symmetry axis. A quantum signature, which is manifested as the breakdown of the spin-induced stabilization, is directly observable by rotating a near-prolate nanodiamond on resonance with the NVs. Resulting from the angular momentum nature of electronic spins and the inherent nonlinearity of three dimensional rigid body rotations, the predicted phenomena open the door towards torque-free quantum coherent interactions between spins and rotation.  This might facilitate the generation and read-out of nonclassical rotation states solely via embedded spins. For instance, using the quantum control methods developed for NV states~\cite{doherty2013nitrogen,perdriat2021spin}, one could switch between  unstable and stable rotations to generate and recombine orientational superpositions via coherent inflation \cite{romero2017coherent}.

\acknowledgements

We thank Klaus Hornberger and Cosimo Rusconi for feedback on the manuscript. Y.M. is supported by the EPSRC center for Doctoral Training on Controlled Quantum Dynamics at Imperial College London (Grant No. EP/L016524/1) and funded by the Imperial College President's PhD Scholarship. M.S.K. acknowledges TheBlinQC funded from the QuantERA ERA-NET Cofund in Quantum Technologies implemented within the European Union's Horizon 2020 Programme and the Korea Institute of Science and Technology (KIST) Open Research Programme. B.A.S. is supported by Deutsche Forschungsgemeinschaft (DFG, German Research Foundation) - 411042854.

\appendix

\section{Effective Hamiltonian}\label{sec:Ham}

We use Euler angles in the $z$-$y'$-$z''$ convention, where the body-fixed principal axes $\mathbf{n}_1$, $\mathbf{n}_2$, $\mathbf{n}_3$ are related to the space-fixed axes $\mathbf{e}_x$, $\mathbf{e}_y$, $\mathbf{e}_z$ by three rotations: first by $\alpha$ around $\mathbf{e}_z$, then by $\beta$ around the new $\mathbf{e}_y$ axis, and finally by $\gamma$ around the new ${\bf e}_z$ axis. Choosing $\mathbf{n}_3$ as the symmetry axis of the particle, $\mathbf{n}_2$ as the spin quantization axis, and the space frame $\mathbf{e}_z$ axis along the direction of $\mathbf{J}$, the total angular momentum along the rotor principal axes is
\begin{subequations}\label{eq:angmom0}
\begin{align}
    J_1=&-J\sin\beta\cos\gamma,\\
    J_2=&J\sin\beta\sin\gamma,\\
    J_3=&J\cos\beta.
\end{align}
\end{subequations}
Also, $J_i=I_i\omega_i+S_i$ where the mechanical rotation rates $\omega_i$ are given by
\begin{subequations}\label{eq:angmom1}
\begin{align}
    \omega_1=&-\dot{\alpha}\sin\beta\cos\gamma+\dot{\beta}\sin\gamma,\\
    \omega_2=&\dot{\alpha}\sin\beta\sin\gamma+\dot{\beta}\cos\gamma,\\
    \omega_3=&\dot{\alpha}\cos\beta+\dot{\gamma}.
\end{align}
\end{subequations}
For $I_1=I_2\equiv I$ and $S_1=S_3=0$, these equations can be combined to yield 
\begin{subequations}
\begin{align}
    \dot{\alpha}= &\frac{J}{I}-\frac{S_2\sin\gamma}{I\sin\beta},\label{eq:alpha}\\
    \dot{\beta}= &-\frac{S_2}{I}\cos\gamma,\label{eq:beta}\\
    \dot{\gamma}= &\cos\beta\frac{I-I_3}{II_3}J+\frac{\cos\beta\sin\gamma}{\sin\beta}\frac{S_2}{I}.\label{eq:gamma}
\end{align}
\end{subequations}

The presence of a spin angular momentum $|S_2|\ll J$ creates an effective potential for $\gamma$. To see this, we take the time derivative of Eq.~\eqref{eq:gamma} and use that for a rapid initial rotation around the spin-quantization axis, $J_2(0) \simeq J$, the symmetry of the rotor guarantees that $\beta\simeq \pi/2$ holds throughout the dynamics. The equation of motion for $\gamma$ is thus approximately
\begin{equation}
    \ddot{\gamma}\simeq \frac{I-I_3}{I^2I_3}JS_2\cos\gamma,
\end{equation}
resulting in the effective Hamiltonian Eq.\,\eqref{eq:heff} once we define the effective moment of inertia as $I_{\mathrm{eff}}=II_3/(I-I_3)$.

For a near-symmetric rotor, $I_1\gtrsim I_2>I_3$ or $I_1\lesssim I_2<I_3$, rapid mid axis rotation still implies $\beta\simeq\pi/2$, yielding
\begin{equation}
    \ddot{\gamma}\simeq -\frac{(I_1-I_2)(I_1-I_3)}{I_1^2I_2I_3}J^2\sin\gamma\cos\gamma+ \frac{I_1-I_3}{I_1I_2I_3}JS_2\cos\gamma,
\end{equation}
which is consistent with the relation given before Eq.\,\eqref{eq:asymS}.

\section{Hard magnet limit}\label{sec:rwa}

If the mechanical rotation frequency $J/I$ is much smaller than $D$ but much larger than $N\hbar/I_k$, the spin dynamics can be adiabatically eliminated (rotating-wave approximation) by transforming into a frame co-rotating with the zero field splitting. Neglecting terms that oscillate with frequency $D$ yields,
\begin{equation}\label{eq:NVhamiltonianMid}
    \hat{H}_{\mathrm{app}}=\sum_{k = 1}^3\left[\frac{\hat{J}_k^2}{2I_k}-\frac{\hat{J}_k}{I_k}\sum_{m = 1}^N(\mathbf{n}^{(m)}\cdot\mathbf{n}_k)(\mathbf{n}^{(m)}\cdot\hat{\mathbf{S}}^{(m)})\right],
\end{equation}
where we dropped contributions $\hat{S}_k^{(m)}\hat{S}_k^{(n)}/I_k$ since $S \ll J$. For each collective NV eigenstate given by a product of $\mathbf{n}^{(m)}\cdot\hat{\mathbf{S}}^{(m)}$ eigenstates with eigenvalues $s^{(m)}=0,\pm1$, the particle rotation is described by the effective hard-magnet Hamiltonian
\begin{equation}
    \hat{H}_{\mathrm{eff}}=\sum_{k = 1}^3\frac{1}{2I_k}(\hat{J}_k-\tilde{S}_k)^2.
    \label{eq:Heff}
\end{equation}
Here, $\tilde{S}_k=\sum_{m = 1}^N(\mathbf{n}^{(m)}\cdot\mathbf{n}_k)\hbar s^{(m)}$. The Heisenberg equations for $\hat{J}_k$ are the Euler equations Eq.~\eqref{eq:euler} in the classical limit.

\begin{figure}[t]
\centering
\includegraphics[width=0.45\textwidth]{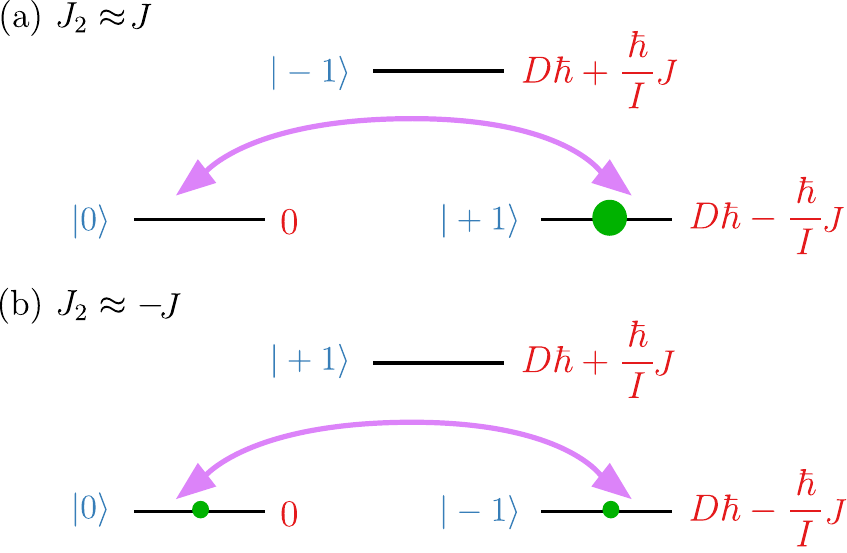}
\caption{Approximate eigenstates (blue) and eigenenergies (red) of the semiclassical Hamiltonian \eqref{eq:semiham} for $J_1, J_3 \ll J$ as well as (a) $J_2 \simeq J$ and (b) $J_2 \simeq -J$. The purple arrows indicate possible spin transitions. The green dot in (a) indicates the initial NV state and in (b) the occupied states after the adiabatic state change.} \label{fig:eng2}
\end{figure}

\section{Semiclassical Dynamics on Resonance}\label{sec:onNV}

The spin dynamics for rapid rotations can be described by a semiclassical model, describing the quantum spin dynamics for parametrically rotating objects. The resulting effective NV Hamiltonian is
\begin{equation}\label{eq:semiham}
    \hat{H}_{\mathrm{semi}}=-\frac{1}{I}J_1(t)\hat{S}_1-\frac{1}{I}J_2(t)\hat{S}_2-\frac{1}{I_3}J_3(t)\hat{S}_3+\frac{1}{\hbar}D\hat{S}_2^2,
\end{equation}
where $J_i(t)$ are the classical angular momentum trajectories. Expressing the spin-1 operators $S_i$ in matrix form gives
\begin{widetext}
\begin{equation}
    \hat{H}_{\mathrm{semi}}=\\
    \begin{pmatrix}
D\hbar-\frac{\hbar}{I}J_2(t) & \frac{i}{\sqrt{2}}\frac{\hbar}{I}J_1(t)-\frac{1}{\sqrt{2}}\frac{\hbar}{I_3}J_3(t) & 0\\
-\frac{i}{\sqrt{2}}\frac{\hbar}{I}J_1(t)-\frac{1}{\sqrt{2}}\frac{\hbar}{I_3}J_3(t) & 0 & \frac{i}{\sqrt{2}}\frac{\hbar}{I}J_1(t)-\frac{1}{\sqrt{2}}\frac{\hbar}{I_3}J_3(t)\\
0 & -\frac{1}{\sqrt{2}}\frac{i\hbar}{I}J_1(t)-\frac{1}{\sqrt{2}}\frac{\hbar}{I_3}J_3(t) & D\hbar+\frac{\hbar}{I}J_2(t)
\end{pmatrix}.
\end{equation}
\end{widetext}
The eigenstates of the NV center zero field splitting $D\hat{S}_2^2/\hbar$ are $|+1\rangle=(1, 0, 0)$, $|0\rangle=(0, 1, 0)$ and $|-1\rangle=(0, 0, 1)$.

For a prolate rotor $(I_3<I)$ off-resonance, $J_2(0)\simeq J\ll ID$, the stabilization of the $\mathbf{n}_2$ axis requires the NV state to be $|+1\rangle = (1,0,0)$, which is an approximate eigenstate of \eqref{eq:semiham} with eigenvalue $D\hbar-\hbar J/I$. Transitions to other states are forbidden due to the large energy differences and the small value of the off diagonal elements. The hard magnet regime is thus valid throughout the dynamics.

If the rotation is fast enough, $J\simeq ID$, the energies of the states $|+1\rangle$ and $|0\rangle = (0,1,0)$ become nearly degenerate, inducing nonadiabatic transitions between the two states [Fig.~\ref{fig:eng2}(a)].  This decreases $\langle \hat{S}_2 \rangle$, rendering the rotations unstable due to Eq.\,\eqref{eq:sym}. This in turn decreases $J_2$, bringing the NV out of resonance so that the spin follows the rotor adiabatically in an eigenstate of \eqref{eq:semiham}. As $J_2$ eventually approaches $-J$, the NV state has a finite $|-1\rangle = (0,0,1)$ component [Fig.~\ref{fig:eng2}(b)], which can transit nonadiabatically to the $|0\rangle$ state. The nonadiabatic evolution is characterized by a rapid change of $\langle\hat{S}_2\rangle$ with relatively constant $\langle\hat{J}_2\rangle$ close to $\pm J$ [see Fig.~\ref{fig:diamond}(b)], while $\langle\hat{S}_2\rangle$ follows $\langle\hat{J}_2\rangle$ during the adiabatic evolution.

For an oblate rotor $(I_3>I)$, stabilization of $J_2(0)\simeq J$ requires the initial NV state to be $|-1\rangle$. This is an approximate eigenstate of energy $D\hbar+\hbar J/I$, which is always detuned from the other states, precluding nonadiabatic transitions.

\end{document}